# Optimization of a Non-arsenic Iron-based Superconductor for Wire Fabrication

J. E. Mitchell[Δ], D. A. Hillesheim[°δ], C. A. Bridges[°], M. P. Paranthaman[°]
K. Gofryk[Δ#], M. Rindfleisch[◊], M. Tomsic[◊], A. S. Sefat[Δ*]

[Δ]*Materials Science & Technology Division, Oak Ridge National Laboratory, Oak Ridge, TN 37831, USA*

[°]*Chemical Sciences Division, Oak Ridge National Laboratory, Oak Ridge, TN 37831, USA*

[◊]*Hyper Tech Research, Inc., Columbus, OH 43228, USA*

[*] *Corresponding author: sefata@ornl.gov*
[#] *Present address: Idaho National Laboratory, Idaho Falls, Idaho 83415, USA*

**Abstract**

We report on the optimization of synthesis of iron selenide-based superconducting powders and the fabrication of selenide-based wire. The powders were synthesized by an ammonothermal method, whereby Ba is intercalated between FeSe layers to produce $Ba_x(NH_3)_yFe_2Se_2$, with tetragonal structure similar to $AFe_2X_2$ ($X$: As, Se), '122', superconductors. The optimal $T_c$ (up to 38 K), Meissner and shielding superconducting fractions, and critical current density ($J_c > 10^5$ $A/cm^2$; 4 K, self-field) are obtained from the shortest reaction time ($t$) of reactants in liquid ammonia (30 minutes). With the increase of $t$, a second crystalline 122 phase, with a smaller unit cell, emerges. A small amount of $NH_3$ is released from the structure above ~ 200 °C, which results in loss of superconductivity. However, in the confined space of niobium/Monel tubing, results indicate there is enough pressure for some of $NH_3$ to remain in the crystal lattice, and thermal annealing can be performed at temperatures of up to 780 °C, increasing wire density and yielded a reasonable $T_c \approx 16$ K. We report of the first successful fabrication of non-arsenic high-$T_c$ 122 iron-based superconductor wires.

**Introduction**

A family of iron-based superconductors adopts crystal structures of *I*4/*mmm* ThCr$_2$Si$_2$-type tetragonal framework, with '122' chemical formula. This family is based on *A*Fe$_2$*X*$_2$ (*A* = alkali or alkaline-earth metals; *X* = Se, As) [1-4]. The structures are composed of edge-shared Fe*X*$_4$ tetrahedral layers separated by *A* [4]. The iron selenium (non-arsenide) materials (FS), have the second largest superconductivity critical temperature ($T_c$ = 30 to 45 K) [1-8] after the rare-earth (*R*) '1111' family of *R*FeAsO ($T_c$ = 56 K). The 122 family is considered as a good candidate for high-field applications because of its high critical field, $H_{c2}$ [9], irreversibility fields ($H_{irr}$) [10], and $J_c$ values [11, 12], in combination with its weak field-dependence on $J_c$ [13, 14]. It is also more adaptable for technological use because it can be made with 100% reduction in critical resistance (*R*) [4], has sharper resistive transitions [15] and lower electronic anisotropy ($\gamma_H \approx 1$ to 2) than other iron-based superconductors [16, 17]. Moreover, 122 offers intrinsic pinning associated with atomic-scale defects resulting from chemical doping [15, 17] and has a large $J_c$ up to 5 × 10$^6$ A/cm$^2$ (2 K) in Ba$_{0.6}$K$_{0.4}$Fe$_2$As$_2$ single crystals [11] and at 4.2 K in BaFe$_{1.8}$Co$_{0.2}$As$_2$ films [14]. In addition, the behavior of $J_c$(H) can be approximately constant at 20 to 30 K, with $J_c$(8 Tesla) remaining high at ~ 10$^5$ A/cm$^2$ in Ba$_{0.6}$K$_{0.4}$Fe$_2$As$_2$ (Fig. 1a) and Ca$_{0.75}$K$_{0.25}$Fe$_2$As$_2$ (Fig. 2) single crystals [18]. In addition, 122 can be completely free of toxic arsenic in *A*Fe$_2$Se$_2$ (FS) [1-3, 4], which is the material of focus of this manuscript.

The FS that are prepared by high-temperature synthesis techniques (e.g. Bridgman) [19-21] are typically referred to as '245' (instead of 122) due to their iron-vacancy disordered structure (e.g. in $A_x$Fe$_{2-y}$Se$_2$) [22]; their superconducting properties are sparse with evidence of non-pure products and phase separations [22-27]. For FS that are synthesized by a low-temperature ammonothermal route, higher purity products are evident with bulk superconducting behavior [28, 29]. Although this solution method allows for greater ease of intercalation of larger alkaline-earth metals and rare-earths (for *A*) between FeSe layers [28, 29], non-deliberate intercalation of ammonia is likely within the spacer layer, producing a Fe-vacancy-free phase of $A_x$(NH$_3$)$_y$Fe$_2$Se$_2$ (we call "N122") [30]. For N122, the superconducting properties may be controlled by electronic doping and lattice expansion, which in turn modify both $T_c$ and the amount of superconducting shielding fraction [30]. Although the in-plane lattice parameters of 245 and N122 are comparable within a narrow range (3.78 to 3.96 Å), the *c*-lattice parameter for N122 is expanded compared to 245, which offers evidence of co-intercalation of larger spacer molecules (such as NH$_3$) together with *A*. For example, the structure of K$_x$Fe$_2$Se$_2$ produced by the high-temperature method is reported with *I*4/*mmm* and $c \approx 14.04$ Å [1] while that synthesized from ammonothermal reaction was indexed using two larger unit cells with $c \approx 16.16$ Å and 20.48 Å [29]; the latter inhomogeneous product may be due to the diffusion-controlled intercalation process. Evidence for N122 co-intercalation by ammonia is additionally offered by neutron diffraction experiments for *A* = Li [30], and is also seen in the resulting smaller *c*-lattice parameter through de-intercalation by heating to ~ 200 °C [31].

In an effort to understand and optimize superconductivity in FS, here we focus on *A* = Ba in *A*Fe$_2$Se$_2$ because it has negligible air-sensitivity (compared to alkali metals such as K), has the highest reported $T_c$ value among FS, and promises reliable bulk superconductivity [29]. We have systematically explored the impact of variations in ammonothermal synthesis conditions on the structure and properties, and in particular on the possible role that intercalated $NH_3$ molecules play in the development of superconductivity for $Ba_x(NH_3)_yFe_2Se_2$. Due to its lower toxicity compared with arsenic iron-based superconductors (e.g. $Ba_{0.6}K_{0.4}Fe_2As_2$) and higher $T_c$ compared with 11 selenides (i.e. $FeTe_{1-x}Se_x$ with $T_c \approx 15$ K) we make some conclusions about this material's potential for wire applications by drawing wire while maintaining a critical current.

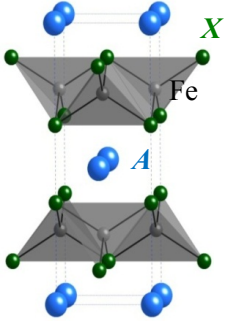

**Figure 1**: (color online) Depiction of crystal structure of $AFe_2X_2$ ($A$ = alkali or alkaline-earth metals; $X$ = Se, As), with $ThCr_2Si_2$-type tetragonal crystal structure adopting $I/4mmm$ space group. For $Ba_x(NH_3)_yFe_2Se_2$, it is found that $NH_3$ occupy the $2b$ sites, while Ba atoms sit at the $2a$ sites [28, 37].

**Experimental**

For the synthesis of N122 $BaFe_2Se_2$, first the tetragonal $\beta$-FeSe had to be synthesized from iron granules (Alfa Aesar, 99.98%) and selenium shots (Alfa Aesar, 99.999+ %) that were sealed in double-walled evacuated fused silica ampoules. The ampoules were heated to 700 °C and held for three days, then the temperature was raised to 1065 °C over three days to form a melt, followed by annealing at 420 °C for three days, after which the product was quenched in an ice-bath. Grinding the slate-grey powder increased the proportion of unwanted hexagonal $\delta$-FeSe or $Fe_7Se_8$, which was subsequently substantially reduced by annealing the powder in vacuum at 420 °C for several more days. Intercalation of small pieces of dendritic barium (Alfa Aesar, 99.9 %) into $\beta$-FeSe powder proceeded by loading stoichiometric amounts of each material in a 100 mL Schlenk tube, in the ratio of Ba:FeSe = 0.5:2 (or in some samples 0.55:2), in a helium-filled glove box with <1 ppm of $O_2$ and $H_2O$. Using a vacuum-gas manifold, approximately 20 to 25 mL of Na-dried liquid ammonia was condensed onto the reactants at liquid-nitrogen temperatures. The $NH_3$ was then allowed to gradually warm to room temperature in the sealed Schlenk vessel, with magnetic stirring begun once the ammonia had liquefied. (Warning: vapor pressures in the range of 8-10 bar are obtained near room temperature over liquid ammonia) The reaction was complete once the deep blue color of the 'solvated electron' had disappeared. The products were kept in liquid $NH_3$ for a total of $t \approx 30$ min (S30m), 5 hours (S5h), or 3 days (S3d); please note that these notations will be used to denote these samples for the rest of this manuscript. After these reaction times, the excess ammonia was condensed into a second Schlenk vessel to isolate the product. The resulting black powder was collected and stored in a helium-filled glove box, with care taken to avoid possible reaction in air. Although the products are not air-sensitive in the short term, a gradual decline in crystallographic quality was observed in air over several days.

Powder X-ray diffraction was performed using a PANalytical X'Pert PRO MPD employing a Ni-filtered Cu-$K_\alpha$ radiation source. The data were analyzed by Rietveld refinement using GSAS-EXPGUI [32, 33]. $Ba_x(NH_3)_yFe_2Se_2$ was indexed to the $ThCr_2Si_2$-type crystal structure with $I4/mmm$ space group. The barium content, $x$, was refined using 0.5 as the starting point, while attempts to include $NH_3$ in the structure provided minimal improvement in agreement indices, and was finally excluded. DC magnetization measurements proceeded using a Quantum Design Magnetic Properties Measurement System. Zero-field cooled (ZFC) and field-cooled (FC) magnetization measurements were performed at 10 Oe from 5 to 50 K; since the magnetization varied linearly with applied field below 100 Oe at 1.8 K noting full shielded response for a superconductor, magnetic susceptibility is presented as $\chi = -1/(4\pi)$. ZFC magnetization data were also collected at 1 kOe from 5 to 300 K (not shown), and did not follow a Curie-Weiss law behavior. M vs. H magnetization loops at 4 K were measured upon increasing field to 6.8 Tesla, decreasing to -6.8 Tesla, then up to 6.8 Tesla, for the estimations of $J_c$. Mass loss measurements were performed using a Perkin Elmer Pyris1 Thermogravimetric Analysis (TGA) instrument located inside a helium-filled glove box with <1 ppm $O_2$ and $H_2O$; ~20 mg of $Ba_x(NH_3)_yFe_2Se_2$ powder were loaded into a ceramic pan with heating at a rate of 40 °C/min from 40 to 800 °C under a 20 L/min flow of He. A portion of the evolved gasses was directed through a fused silica capillary into a Hewlett-Packard 5970 Quadruple Mass Selective Detector (MS); molecular masses were scanned at a rate of 1.63 scans/s from 5 to 500 AMU. A significant atmospheric $N_2$ baseline arises in the mass spectrograph due to air leaking backward from the exhaust pump, which does not interfere with the interpretation of the results.

For wires, the powders were packed into niobium tubing that is surrounded by Monel, and then plugged at both ends in a helium-filled glove box with less than 1 ppm levels of $O_2$ and $H_2O$. The tube was drawn out into a wire of 0.83 mm diameter at Hyper Tech Research Inc.. The total length of wire produced was ~ 41 feet. The as-drawn wire was found to be non-superconducting, and a post-annealing step of both as-drawn and rolled wires at 780 °C for 3 days was crucial in reviving the superconducting nature of FS material. For the annealing process, ~10 cm middle section of the 41 ft. wire was cut and crimped at both ends inside the glove box, then sealed in a 1/3 filled argon atmosphere silica glass, followed by thermally-annealing in a furnace. The cross-sectional wire samples were prepared by standard metallographic techniques. The optical images of these wires are shown in Fig. 2. As shown in Fig. 2, there is a clear interaction between the superconductor and the niobium core during post-annealing. However, it is still possible there is an equilibrium between $NH_3 \leftrightarrow N_2$ and $H_2$, with enough pressure for some of $NH_3$ to stay. The electrical resistivity and critical current were measured on ~ 1 cm long (cut and crimped inside the glove box) pieces, measured in a commercial system (PPMS-14, Quantum Design) using a four-probe method. Electrical contacts were made by spot-welding platinum wires to the specimens.

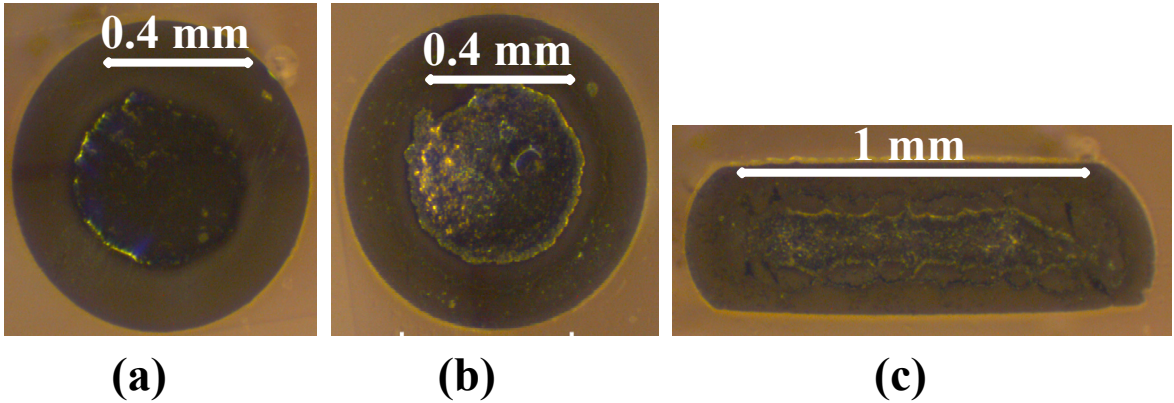

**Figure 2**. The optical cross-sectional images of $Ba_x(NH_3)_yFe_2Se_2$ wires of : (a) as-drawn; (b) post-annealed at 780 °C for 3 days; and (c) rolled and annealed at 780 °C for 3 days.

**Results**

The powder X-ray diffraction patterns for the products S30m, S5h, and S3d are shown in Fig. 3a. In each case, two phases adopting the $ThCr_2Si_2$-type tetragonal structures were observed, differing primarily in the length of the *c*-lattice parameter. The larger unit cell (LC) generally had a *c*-parameter of ~ 16.7 Å, while the smaller unit cell (SC) adopted a *c*-axis length near 15.6 Å (Table 1). The relative proportion of SC increased with increasing *t*, i.e. S30m gave LC:SC ≈ 76:10 by mass, while S3d showed LC:SC ≈ 10:77. Table 1 shows that the Ba-Se bond length in both tetragonal phases is longer than would be expected for direct bonding to occur (~ 3.4 Å) [34]; this may suggest that $NH_3$ co-intercalates along with barium to push the FeSe layers further apart, or at least that the interactions are weak. In addition to 122 tetragonal LC and SC phases, several impurity phases were observed, including $Fe_7Se_8$ and both orthorhombic and monoclinic phases of $Ba(OH)_2 \cdot H_2O$. Although efforts were made to use shiny, untarnished silvery Ba pieces, the presence of barium hydroxides may have originated from contaminated Ba surfaces, or may be the result from the reaction of residual, loose, Ba with air-moisture during X-ray diffraction data collection. Table 1 summarizes the results, while Fig. 3b shows the trend in the size of lattice parameters *c*/*a* with decreasing temperature; no change in the linear trends was observed around $T_c$.

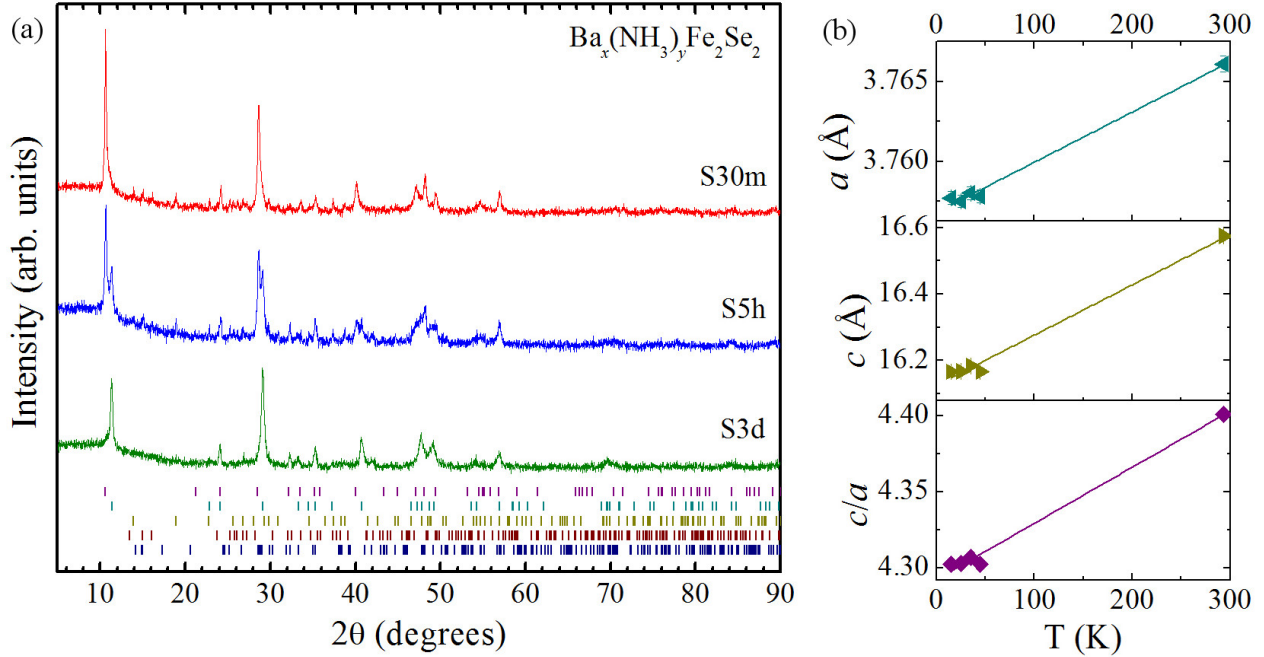

**Figure 3**. (color online) (a) X-ray powder diffraction patterns for samples S30m (top), S5h (middle), and S3d (bottom); Tick marks at the bottom indicate the diffraction peaks for phases, from top to bottom: LC N122, SC N122, orthorhombic $Ba(OH)_2 \cdot H_2O$, monoclinic $Ba(OH)_2 \cdot H_2O$, and $Fe_7Se_8$. (b) Comparison of lattice parameters, $a$, $c$, and $c/a$ at room temperature and below 45 K, showing the trend in shrinking lattice upon cooling.

The temperature dependence of normalized magnetic susceptibility, $4\pi\chi(T)$, on S30m, S5h, and S3d are shown in Figures 4a, c, and e, respectively. The $T_c$ values are in the range of 35 to 38 K (taken as the divergence of ZFC and FC data), with higher $T_c$ noted for shorter reaction times. Similarly, the shielding and Meissner fractions decrease with increasing $t$. For example, the superconducting fraction drops from 76 % for S30m to ~20 % for S3d, and the Meissner fraction declines from 14 % to ~3 %, respectively. The $\chi(T)$ measurements were done on multiple samples from each of the batches, and were found to be reproducible. Additionally, S3d demonstrates a positive offset of FC data, suggesting the development of paramagnetic domains through the prolonged contact with liquid $NH_3$. Field dependence of magnetization, M(H), is shown in Figures 4b, d, and f for two quadrants only. No upper critical field is observed up to 6.8 Tesla. In order to estimate the critical current density at 4 K, we used Bean's formula; $J_c = 30\Delta M/<R>$, where $\Delta M$ is the difference in magnetization at 6 Tesla, and $<R>$ is the average particle size estimated from scanning electron micrographs (SEM) of the powders, which are estimated to be typically ~ 10 µm. The $J_c$ values are listed in Table 1.

**Table 1**. Structural data, $T_c$, and $J_c$ estimates for $Ba_x(NH_3)_yFe_2Se_2$. Ammonia molecules were omitted from the refinement, i.e. only the $Ba_xFe_2Se_2$ 122 lattice was considered. For the refined structural parameters, the LC phase is shown on the top line, with the SC phase on the bottom.

| Sample name | Rxn t | x Nom. | x refined | a (Å) | c (Å) | Ba-Se length (Å) | Mass % | $T_c$ (K) | $J_c$ (×10⁵ A/cm²) at 4K |
|---|---|---|---|---|---|---|---|---|---|
| S30m | 30 min | 0.55 | 0.444(4) | 3.7728(4) | 16.674(2) | 3.722(4) | 75.7(2) | 38 | 3.6 |
|   |   |   | 0.50(2) | 3.785(1) | 15.98(2) | 3.62(2) | 9.9(3) |   |   |
| S5h | 5 hours | 0.5 | 0.383(5) | 3.7798(5) | 16.691(3) | 3.783(5) | 47.4(3) | 36 | 3.9 |
|   |   |   | 0.395(7) | 3.8021(8) | 15.580(6) | 3.578(7) | 34.6(3) |   |   |
| S3d | 3 days | 0.5 | 0.442(2) | 3.794(1) | 16.20(1) | 3.72(2) | 9.9(4) | 35 | 2.6 |
|   |   |   | 0.480(4) | 3.8005(4) | 15.499(3) | 3.557(3) | 76.9(1) |   |   |

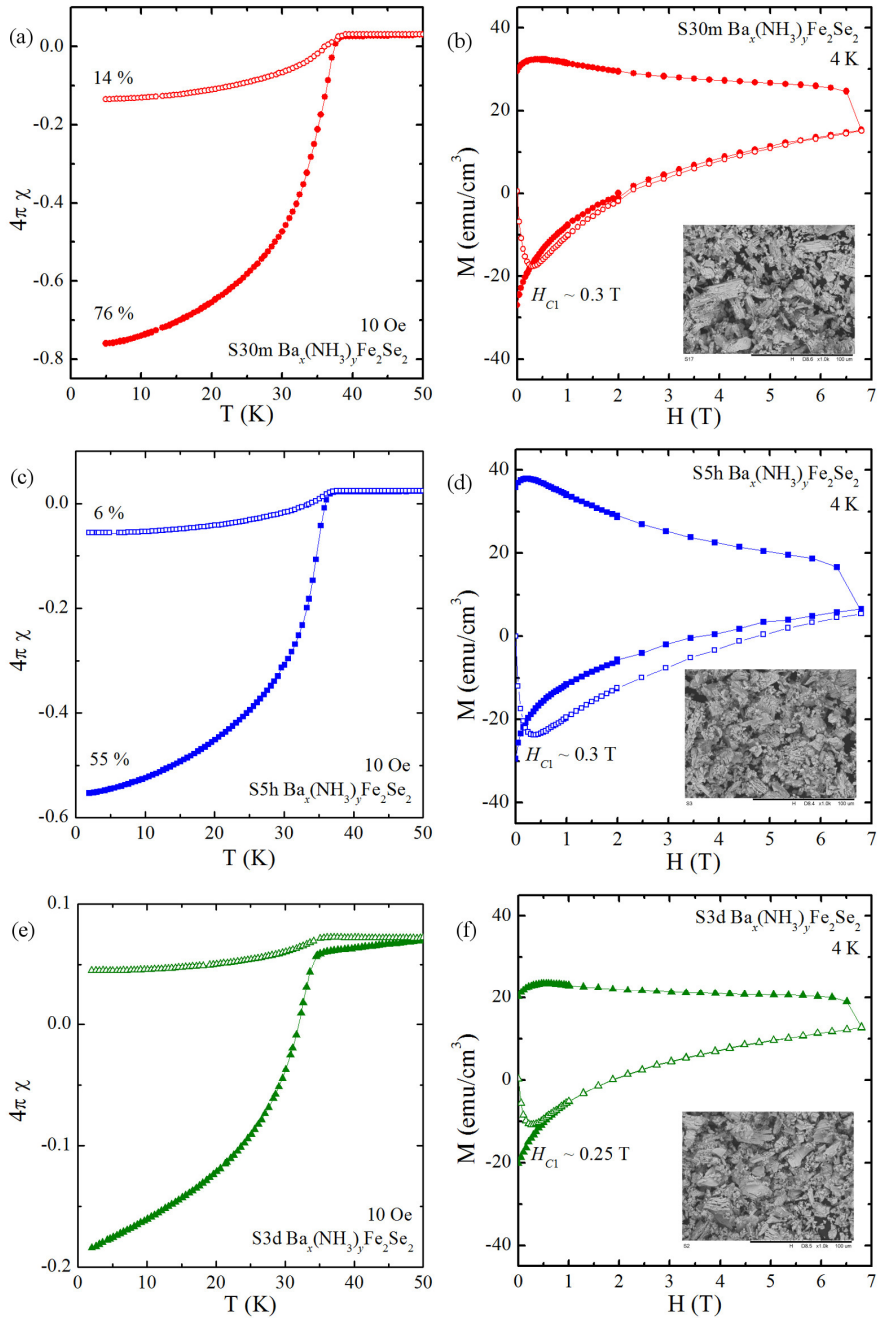

**Figure 4**. (color online) Magnetization data for $Ba_x(NH_3)_yFe_2Se_2$ for samples S30m (a, b), S5h (c, d), and S3d (e, f). Temperature dependence of magnetization curves in 10 Oe field from 2 to 50 K showing $T_c$ = 35 to 38 K (a, c, e). Field dependence of magnetization at 4 K show approximately similar behavior (b, d, f); Insets are SEM images of the powder that are used to estimate the average particle size of ~ 10 µm.

The TGA-MS results for S30m, S5h, and S3d are shown in Figure 5. Mass loss for $Ba_x(NH_3)_yFe_2Se_2$ proceeded in either two or three steps. All plots show $T_{onset,max}$ ~ 370 °C and $T_{onset,med}$ ~ 280 °C, while S30m and S3d also show a small $T_{onset,min}$ ~ 150 to 200 °C, which is not separately distinguished in the S5h plot. Nonetheless, the combined mass loss at $T_{onset,med}$ and $T_{onset,min}$ is similar for all three samples ($\Delta m_{>280 °C}$ ~ 0.6 to 0.9 % total mass), corresponding to ~ 0.12 to 0.19 $NH_3$ molecules per formula unit, assuming a composition of $Ba_{0.5}Fe_2Se_2$ for the remaining elements. The mass loss related to $T_{onset,max}$ shows much greater variation, ranging from $\Delta m_{370 °C}$ ~ 2.4 % for S3d to $\Delta m_{370 °C}$ ~ 3.6 % for S30m, corresponding to a loss of 0.50 to 0.75 $NH_3$ molecules per "$Ba_{0.5}Fe_2Se_2$", respectively. These values can be combined to derive $y$ = 0.66 for S3d to $y$ = 0.87 for S30m. As previously noted, the mass spectrographs include a large $N_2$ background signal due to an atmospheric back-leak from the MS exhaust pump; however, the peaks corresponding to mass loss in the TGA were derived entirely from $NH_3$.

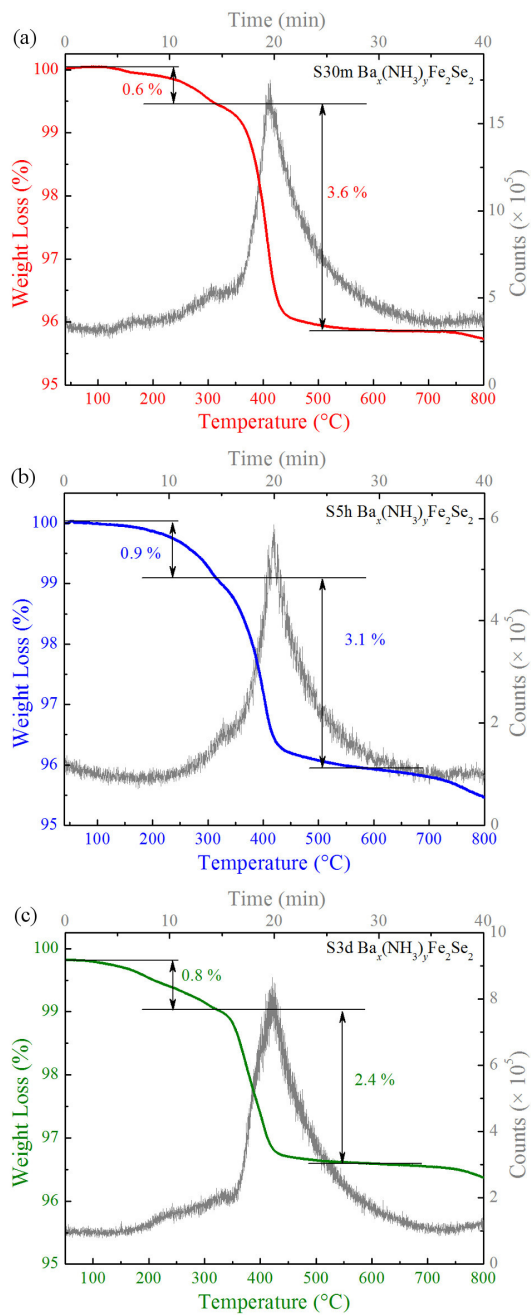

**Figure 5**. TGA-MS results for (a) S30m, (b) S5h, and (c) S3d. TGA mass loss is shown as a solid colored line while total ion count from mass spectrometry is presented in grey. The peaks can be attributed entirely to $NH_3$.

Several annealing studies were performed using sample S5h. $Ba_x(NH_3)_yFe_2Se_2$ powder was pressed into pellets and sealed in fused silica tubes under various atmospheres (vacuum, 0.3 to 0.9 atm $NH_3$ gas, 0.3 atm Ar gas). These were then heated to between 150 °C and 200 °C for 24 hours. The results for the sample post-annealed in Ar at 200 °C shown in Fig. 6 are typical of what was observed under all annealing conditions. Structurally, the LC phase disappears completely, leaving only the SC phase and an increased proportion of $\beta$-FeSe (Fig. 6a). The magnetic susceptibility (Fig. 6b) shows the complete suppression of superconductivity. The origin of the broad maximum at ~ 100 K is not fully understood, but likely derives from paramagnetic impurities that exhibit short range or lower dimensional correlations.

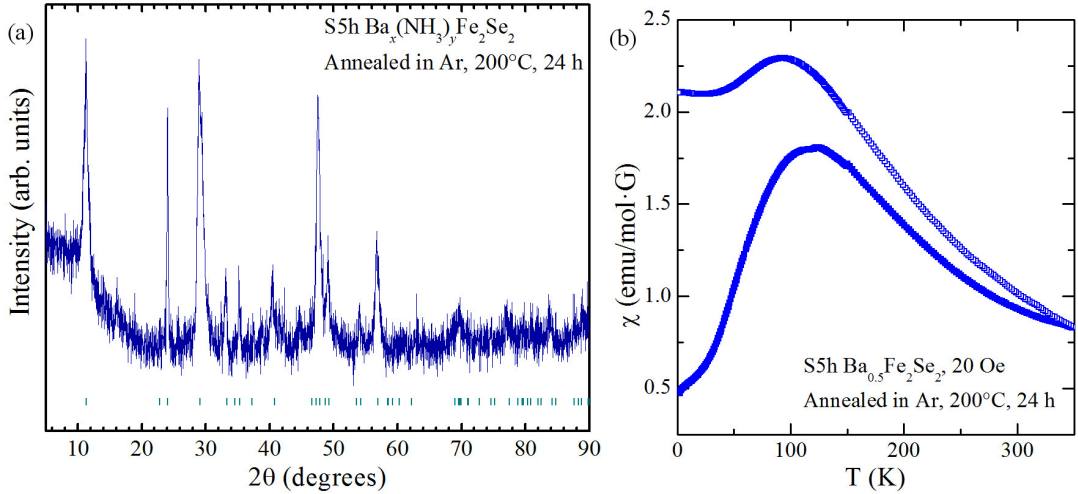

**Figure 6**. (a) Powder x-ray diffraction pattern of S5h annealed at 200 °C in 1/3 atmosphere of argon gas. Tickmarks are Bragg reflection peaks expected for SC phase only. (b) Temperature dependent magnetic susceptibility for S5h annealed at 200 °C in 1/3 atmosphere of argon, showing no diamagnetism.

Figure 7 shows the temperature dependence of normalized electrical resistivity for $Ba_x(NH_3)_yFe_2Se_2$ wires. In general, the $R(T)$ is similar in these two materials, having $T_c \sim 16$ K. In order to achieve a superconducting transition in the wires, thermal annealing of the wires was necessary at 780 °C, as described above, with worse superconducting results at other annealing temperatures. At such high temperatures, it is possible for some of $NH_3$ to leave the crystal lattice that cause structural and $T_c$ degradation, compared with bulk material. A two-step transition observed below $T_c$ could result from a small but noticeable spatial distribution of $T_c$ within the crystals, being slightly smaller for the annealed non-rolled sample. Since Nb metal has a much lower $T_c$ and does not have a continuous conducting path due to its reaction with $Ba_x(NH_3)_yFe_2Se_2$, it does not contribute to the superconducting properties of $Ba_x(NH_3)_yFe_2Se_2$ wires. The presence of better superconducting characteristics in this sample are also highlighted by the critical field value. As shown in the inset of Fig. 7 (a) the critical field of the rolled sample is estimated to be 13 Tesla. For the annealed wire the value of $H_{c2}$ is higher and an analysis using the Werthammer-Helfand-Hohenberg model [35, 36] leads to $H_{c2} \sim 21$ T. Figure 7(b) shows the temperature dependence of critical current of the superconducting wires. The overall shape of $I_c$ is typical of superconducting materials. The higher value of $I_c$ observed for the annealed wire is consistent with the resistivity data. The inset of Fig. 7(b) presents a magnetic field dependence of critical current for these two samples measured at 4 K.

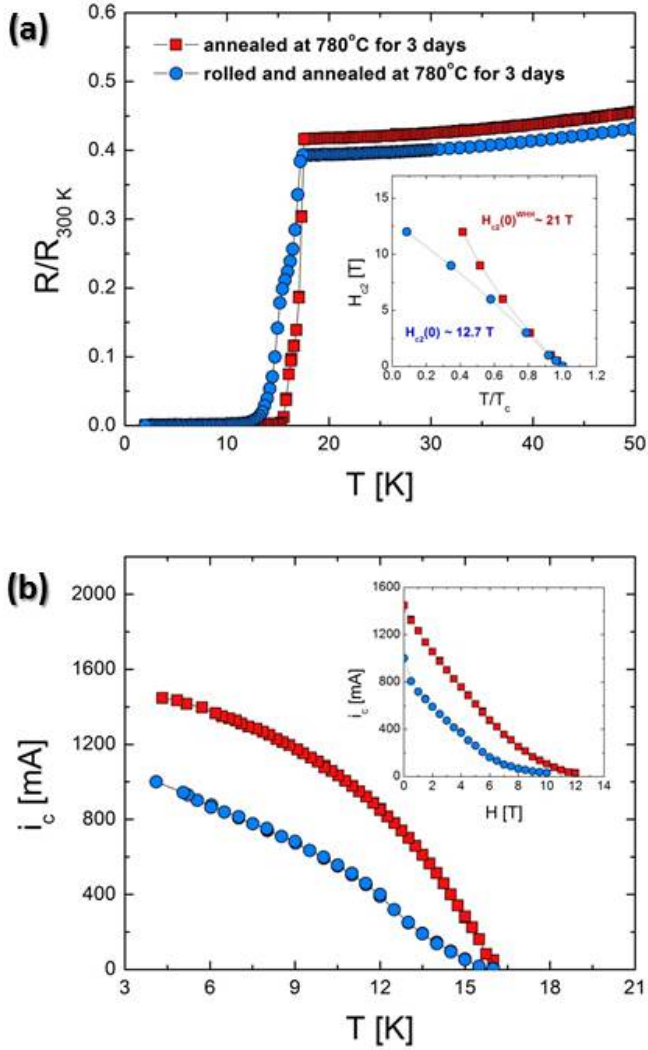

**Figure 7**. (a) The normalized electrical resistivty of $Ba(NH_3)Fe_2Se_2$ superconducting wires. The inset: critial field vs. normalized temperature. (b) The temperature dependence of the critical current. Inset: $i_c$ vs. H measured at 4 K.

**Discussion**

Based on X-ray diffraction patterns, the LC phase initially forms in the ammonothermal reaction of Ba with $\beta$-FeSe, gradually decomposing with increasing reaction time to form the SC phase (Fig. 3). Based on TGA-MS data (Fig. 5), the decrease in LC phase appears to result from a loss of structural $NH_3$, which also correlates with a decrease in superconducting fraction for the materials studied (Fig. 4). These data combine to suggest the existence of LC phase as the primary superconducting phase. This conclusion is supported by annealing studies (Fig. 6), where the complete suppression of superconducting behavior is accompanied by the disappearance of the LC phase in favor of the SC phase. It is likely that the structural variation that increases the unit cell length in the LC phase also plays a role in inducing superconductivity. This result is contrary to the situation discovered for the analogous $K_x(NH_3)_yFe_2Se_2$ system as described by Ying *et al* [31] where the phase with the larger unit cell was found to have a lower $T_c$ than the smaller cell phase. The TGA-MS results (Fig. 5) seem to indicate at least two distinct mechanisms of $NH_3$ intercalation. One possible explanation is that $NH_3$ intercalates at two different sites, with the majority entering "stable" sites and a smaller amount entering less stable sites or on the surface, which release ammonia at lower temperatures. The annealing studies may suggest that occupation of these lower stability $NH_3$ sites leads to the larger unit cell and induction of superconducting behavior. If this were the case, one would expect a larger amount of ammonia to be released at the first step in samples with larger superconducting fractions, i.e. S30m. This is contrary to what is observed, in that S30m actually has the smallest mass loss at lower temperatures of the three samples presented. A more likely explanation is that $NH_3$ occupies only one site, but undergoes partial decomposes to $NH_2^-$, resulting in more than one onset temperature as the amine further decomposes to $N_2$ and $H_2$. In this scenario, the SC phase only requires a certain minimum amount of $NH_3/NH_2^-$ to remain stable, while any 'excess' ammonia expands the unit cell and induces superconductivity. This explanation requires better structural understanding of $Ba_x(NH_3)_yFe_2Se_2$ than can be determined from traditional powder x-ray diffraction techniques. Neutron studies of $Li_x(NH_3)_yFe_2Se_2$ have determined only one $NH_3$ site with disorder in the position of the hydrogen atoms [28], while synchrotron x-ray diffraction has indicated a similar arrangement for the Ba-system [37]. Furthermore, it remains unclear whether ammonia is the only intercalated species, or whether amine ($NH_2^-$) also enters the structure, and whether the chemical distinction between them has any bearing on superconducting properties, a possibility raised by Scheidt *et al* [30]. Further optimization is necessary to improve the superconducting properties of N122 wires.

**Conclusion**

Ammonothermal synthesis of $Ba_x(NH_3)_yFe_2Se_2$ proceeds quickly, with complete decoloration of the solution indicating the endpoint within about 20 minutes. Prolonging exposure of the product to liquid ammonia promotes degradation of the dominant LC phase to the SC phase, which has a shorter *c*-axis. This also affects the superconducting properties, diminishing the superconducting and Meissner fractions, while the longest exposure time also decreases $J_c$. Evidence from TGA-MS and annealing experiments may point to differences in $NH_3$ intercalation between the LC and SC phases, with evidence suggesting that LC phase may be primarily responsible for superconducting behavior. Nonetheless, the nature of these differences and their role inducing superconductivity is not yet clear. Locating the $NH_3$ molecules within the $Ba_x(NH_3)_yFe_2Se_2$ lattice, and particularly identifying the distinctions between the LC and SC phase, may elucidate the relationship between structure and electronics. Property measurements of vacuum post-annealed samples show a loss of superconductivity, indicating the importance of co-intercalated ammonia in stabilizing superconductivity in $Ba_x(NH_3)_yFe_2Se_2$. However, in the confined space of niobium/Monel tubing, thermal annealing at 780 °C maintains superconductivity. We have successfully demonstrated the preparation of superconducting wires of N122 with $H_{c2}$ ~21 T and carrying critical currents. To the best of our knowledge, this is the first report of wire fabrication of non-arsenic 122 iron-based superconductors.

**Reference**
[1] Guo, J., et al. *Phys. Rev. B* **82**, 180520 (2010).
[2] Wang, D. M., et al. *Phys. Rev. B* **83**, 132502 (2011).
[3] Fang, M. H., et al. *Europhys. Lett.* **94**, 27009 (2011).
[4] Sefat, A. S., et al. *Mater. Research Bull.* **36**, 614 (2011).
[5] Rotter, et al., *Phys. Rev. Lett.* **101**, 107007 (2008).
[6] Sasmal, K., et al. *Phys. Rev. Lett.* **101**, 107007 (2008).
[7] Saha, S. R., et al. *Phys. Rev. B* **85**, 24525 (2012).
[8] Gao, Z., et al. *Europhys. Lett.* **95**, 67002 (2011).
[9] P. M. Aswathy, et al. *Appl. Phys. Lett.* 93 (2008), 32503.
[10] Nature Mater. 11 (2012), 682.
[11] H. Yang, et al. *Appl. Phys. Lett.* 93 (2008), 142506.
[12] M. Putti, et al. *Super. Sci. Tech.* 23 (2010), 034003.
[13] X. Wang, et al. *Adv. Mater.* 21 (2009), 236.
[14] S. Lee, et al. *Nature Mater.* 9 (2010), 397.
[15] A. S. Sefat, *Phys. Rev. Lett.* 101, (2008), 117004.
[16] H. Q. Yuan, et al. *Nature* 457, 565.
[17] A. Yamamoto, et al. *Appl. Phys. Lett.* 94 (2009), 062511.
[18] N. Haberkorn, et al. *Phys. Rev. B* 84 (2011), 64533.
[19] Ren, Z. A., et al., *Europhys. Lett.* **83**, 17002 (2008).
[20] Hanna, T., et al. *Phys. Rev. B* **84**, 24521 (2011).
[21] Wang, C., et al. *Europhys. Lett.* **83**, 67006 (2008).
[22] Bacsa, J., et al. *Chem. Sci.* **2**, 1054 (2011).
[23] Texier, Y., et al. *Phys. Rev. Lett.* **108**, 237002 (2012).
[24] Li, W., et al. *Nature Phys.* **8**, 126 (2012).
[25] Ricci, A., et al. *Phys. Rev. B* **84**, 60511 (2011).
[26] Ksenofontov, A., et al. *Phys. Rev. B* **84**, 180508 (2011).
[27] Charnukha, A., et al. *Phys. Rev. Lett.* **109**, 017003 (2012).
[28] Burrard-Lucas, M., et al. *Nature Mater*. **12**, 15 (2013).
[29] Ying, T. P., et al. *Scientific Reports* **2**, 426 (2012).
[30] Scheidt, E. W., et al. *Eur. Phys. J. B* **85**, 279 (2012).
[31] Ying, T., et al., *J. Am. Chem. Soc.* **135**, 2951 (2013).
[32] Larson, A. C. et al. "General structure analysis system (GSAS)", *Los Alamos National Laboratory Report LAUR* 86-748 (2000).
[33] Toby, B. H. EXPGUI, a graphical user interface for GSAS, *J. Appl. Cryst.* **34**, 210 (2001).
[34] Shannon, R. D., *Acta Cryst. A* **32**, 751 (1976).
[35] Helfand E, et al. *Phys. Rev.* **147**, 288 (1966).
[36] Werthamer N. R., et al. *Phys. Rev.* **147**, 295 (1966).
[37] Shibasaki, S., et al. *http://arxiv.org/abs/1306.0979* (2013).

**Acknowledgement**

Iron-based superconductor wire fabrication research was supported by the Oak Ridge National Laboratory (ORNL) SEED funding program. A.S. also acknowledges support by the U. S. Department of Energy, Office of Science, Office of Basic Energy Sciences, Materials Science and Engineering Division, for some of the powder synthesis efforts. The team acknowledges C. Cantoni for group discussions, and also specifically for tube filling, wire cutting, and keeping annealing-temperature records of wires.